\documentclass[%
  reprint,
  superscriptaddress,
  amsmath,
  amssymb,
  aps,
  prl,
  floatfix
]{revtex4-2}
\setcitestyle{showtitles}
\usepackage{graphicx} 
\usepackage{braket}
\usepackage{color}
\usepackage{amsmath, amssymb, dsfont}
\usepackage{hyperref}
\usepackage{physics}
\usepackage{bm}
\usepackage{changes}
\usepackage{textcomp}
\usepackage{float}

\usepackage{hyperref}
\begin{document}

\title{Resonant Excitation Induced Vibronic Mollow Triplets}
\author{Devashish Pandey}
\affiliation{Department of Electrical and Photonics Engineering, Technical University of Denmark, 2800 Kgs. Lyngby, Denmark}
\author{Corné Koks} 
\affiliation{Department of Electrical and Photonics Engineering, Technical University of Denmark, 2800 Kgs. Lyngby, Denmark}
\affiliation{NanoPhoton – Center for Nanophotonics, Technical University of Denmark, 2800 Kgs. Lyngby, Denmark}
\author{Martijn Wubs}
\affiliation{Department of Electrical and Photonics Engineering, Technical University of Denmark, 2800 Kgs. Lyngby, Denmark}
\affiliation{NanoPhoton – Center for Nanophotonics, Technical University of Denmark, 2800 Kgs. Lyngby, Denmark}
\author{Nicolas Stenger} 
\affiliation{Department of Electrical and Photonics Engineering, Technical University of Denmark, 2800 Kgs. Lyngby, Denmark}
\affiliation{NanoPhoton – Center for Nanophotonics, Technical University of Denmark, 2800 Kgs. Lyngby, Denmark}
\author{Jake Iles-Smith}
\affiliation{School of Mathematical and Physical Sciences, The University of Sheffield, Sheffield S10 2TN, United Kingdom}


\begin{abstract}
 The Mollow triplet is the definitive spectral signature of an optically dressed quantum emitter. We predict that for emitters coupled to localized phonons, this signature is not confined to the zero-phonon line. Under a strong resonant drive, we show that Mollow triplets are strikingly replicated on the associated phonon sidebands —a surprising result, given that phonon sidebands are typically viewed as incoherent, inelastic scattering pathways. These \textit{vibronic Mollow triplets} are a direct fingerprint of dynamically generated dressed states that hybridize the emitter's electronic, photonic, and vibrational degrees of freedom. We develop a scalable analytical formalism to model this effect in complex, multi-mode molecular systems, such as dibenzoterrylene. Our work provides the precise driving conditions for observing these novel spectral features, establishing a new signature of coherence in vibronically coupled systems.
\end{abstract}

\maketitle

The Mollow triplet is a cornerstone of quantum optics, signifying the strong coupling regime where a resonant light field dresses a quantum emitter~\cite{cohen1977dressed,cohen2024atom}. This interaction generates hybridized energy levels $\ket{\pm_n} = (\ket{e, n-1} \pm\ket{g,n})/\sqrt{2}$, entangling the electronic and photonic degrees of freedom. The resulting triple-peak spectrum~\cite{Mollow1969} serves as the definitive hallmark of a highly coherent emitter.
Originally identified in atoms~\cite{F_Schuda_1974, hartig_study_1976, grove_77}, the Mollow triplet has become a ubiquitous benchmark in solid-state and molecular quantum technologies, spanning semiconductor quantum dots~\cite{ulhaq_cascaded_2012,Ulhaq2013Detuning-dependentDot, Jesper_Mørk}, defect centers in bulk~\cite{Konthasinghe:19} and low-dimensional materials~\cite{gerard2025res}, and organic molecules~\cite{wrigge2008efficient, Ott}. 

For the Mollow triplet to appear in resonance fluorescence, the Rabi frequency must overcome both the spontaneous emission and decoherence rates. In solid-state emitters, electron-phonon interactions are a dominant decoherence mechanism. For example, the effects of bulk acoustic phonons—extensively studied in quantum dots—are well understood, causing damped Rabi oscillations~\cite{Ramsay2010damping,mccutcheon2011damping}, Rabi frequency renormalization~\cite{Ramsay2010renorm, mccutcheon2010quantum, Wei2014renorm}, and the revival of coherent scattering above saturation~\cite{Mccutcheon2013coh,iles2019vibrational}.
However, many emitters, including defect centers and molecules, also couple to high-quality-factor localized phonon modes~\cite{fischer2023combining}. These modes facilitate the formation of vibronic states~\cite{PhysRevResearch.7.021001}—hybrid electronic-vibrational manifolds that generate spectrally distinct phonon sidebands governed by Franck-Condon physics~\cite{fox2010optical}. Despite the ubiquity of these features, the impact of localized phonon interactions on the Mollow physics remains largely unexplored.

In this Letter, we bridge this gap by predicting a novel phenomenon: the emergence of a Mollow triplet (MT) not only on the resonantly driven zero-phonon line (ZPL) but also on its associated phonon sidebands (PSBs). This observation challenges the conventional view of PSBs as purely incoherent inelastic scattering pathways. We demonstrate that these sideband triplets serve as a direct spectral fingerprint of the drive-induced dressed states imprinted onto the vibronic transitions, revealing a new class of \textit{vibronic dressed states}.

To provide a predictive framework for complex molecular systems, we develop a scalable analytical formalism that rigorously captures the multi-mode vibronic coupling characteristic of realistic solid-state emitters. We apply this model to dibenzoterrylene (DBT), 
a molecular system exhibiting quantum emission with near transform-limited lineshapes at cryogenic temperatures~\cite{Nicolet, Zirkelbach2022High-resolutionCrystal, Clear2020Phonon-InducedMolecules} and strong coupling to multiple discrete phonon modes~\cite{PhysRevResearch.7.021001}. Our analysis provides a concrete experimental guide for observing these vibronic dressed states and establishes the driving conditions required to resolve the PSB triplets, opening a new avenue for coherently manipulating coupled electron-phonon systems.

\begin{figure*}[t]   
  \centering
  \includegraphics[width = \textwidth]{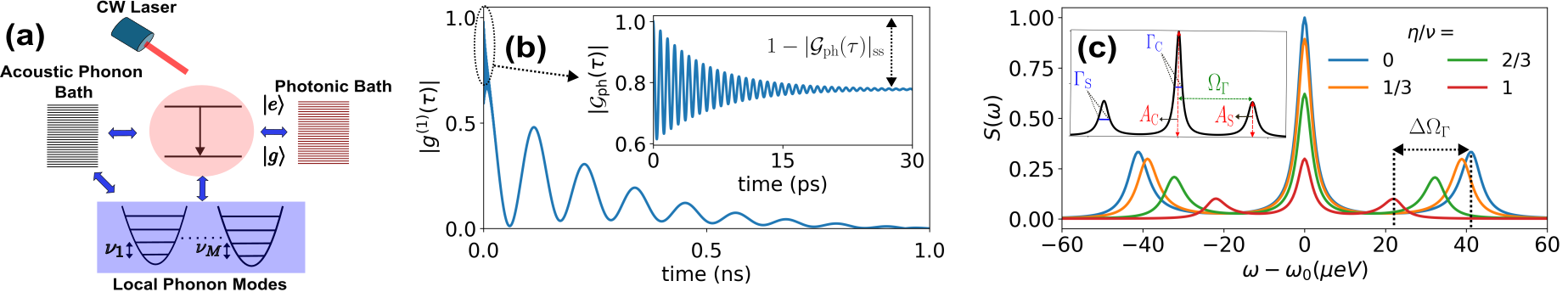}
  \caption{(a) Schematic of a two-level system driven by a CW laser and coupled to local phonon modes, forming an open quantum system that interacts with both photonic and acoustic phonon baths. (b) Coherence decay of an emitter coupled to a local phonon mode with $(\eta/\nu)^2= 0.25$ showing a rapid picosecond-scale drop (Inset) accompanied by visible Rabi oscillations. (c) Comparison of the Mollow triplet for $\eta = 0$ to $\eta = 1$. The parameters used are $\gamma = 4.1\;\mu\mathrm{eV}$, $\Omega = 10\gamma$, $\nu = 5$ meV and $\kappa = 0.2$ meV. (Inset) Mollow spectrum showing the triplet linewidths and amplitudes used to define the linewidth ratio $R_\Gamma = \Gamma_{\rm C}/\Gamma_{\rm S}$ and amplitude ratio $R_{\rm A} = A_{\rm S}/A_{\rm C}$. }
  \label{fig:main}
\end{figure*}

\emph{Model ---} We consider a two-level quantum emitter (TLE) driven by a continuous-wave laser and coupled to $M$ localized phonon modes and a photonic continuum [Fig.~\ref{fig:main}(a)]. Working in a frame rotating at the laser frequency ($\omega_\mathrm{L}$) and within the rotating wave approximation, the total Hamiltonian ($\hbar=1$) is defined as $H = H_{\rm S} + H_{\rm vib} + H_{\rm env}$. The driven electronic system is governed by $H_{\rm S} = (\Delta \sigma_z + \Omega \sigma_x)/2$, with Rabi frequency $\Omega$ and detuning $\Delta = \omega_0 - \omega_\mathrm{L}$. The vibrational contribution, $H_{\rm vib} = H_{\rm ph} + H_{\rm e-ph}$, describes the free phonon evolution and the longitudinal electron-phonon coupling~\cite{mahan2013many}:
$
H_{\rm vib} = \sum_{j= 1}^M \left[ \nu_j b_j^\dagger b_j + \eta_j \sigma^\dagger\sigma (b_j^\dagger + b_j) \right],
$
where $b_j^\dagger$ creates a phonon of frequency $\nu_j$ coupled to the emitter with strength $\eta_j$. Finally, $H_{\rm env}$ describes the continuum photon bath, $\sum_{q}\omega_{q} a_{q}^\dagger a_{q}$, and its transverse dipole interaction with the TLE, $\sum_q (h_q a_{q}^\dagger\sigma + \mathrm{H.c.})$.

Modeling molecular emitters requires a non-perturbative treatment of strong coupling to multiple localized phonon modes. While this can be achieved by extending the system Hamiltonian~\cite{Clear2020Phonon-InducedMolecules}, the resulting Hilbert space scales exponentially with the number of modes, rendering it intractable for complex molecules like DBT. To circumvent this bottleneck, we apply a polaron transformation defined by the unitary $U =\dyad{g} + \dyad{e}B_-$, where $B_\pm = \exp(\pm\sum_{j}\frac{\eta_j}{\nu_j}(b_j^\dagger - b_j))$ is the collective displacement operator. This transformation dresses the electronic states with the vibrational modes of the environment~\cite{Nazir2016ModellingDots}, naturally incorporating the strong electron-phonon correlations into the basis states without expanding the computational overhead.

Within the polaron frame, we construct the master equation in which the electromagnetic environment has been traced out to obtain
\begin{equation}\begin{split}
\frac{d\rho_s(t)}{dt} = &-i[\tilde{H}_s, \rho_s(t)] + \frac{1}{2}\sum_{j= 1}^M\left(\kappa_j^-\mathcal{D}_{b_j}+ \kappa_j^+\mathcal{D}_{b^\dagger_j}\right)\\ 
&+ \frac{\gamma}{2} \mathcal{D}_{\sigma B_-}+ \gamma_{\rm pd}\mathcal{D}_{\sigma^\dagger\sigma},
\label{ME_pl}
\end{split}\end{equation}
where $\mathcal{D}_A\equiv 2A\rho A^\dagger - (A^\dagger A\rho + \rho A^\dagger A )$ is the Lindblad  dissipator. A detailed derivation of this master equation is given in the SI.
The effective system Hamiltonian, $\tilde{H}_s$, describes the coupled emitter-phonon dynamics,
$$
\tilde{H}_s = \tilde{\delta} \sigma^\dagger\sigma + \sum_{j = 1}^M\nu_j b_j^\dagger b_j + \frac{\Omega}{2}(\sigma B_- + \sigma^\dagger B_+).
$$
Here, $\tilde{\delta} = \tilde{\omega}_0 -\omega_l$, where $\tilde{\omega}_0 = \omega_0 - \sum_{j = 1}^M\eta_j^2/\nu_j$ is the polaron-shifted transition and our laser frequency $\omega_l$ is resonant to this transition ($\ket{g,0}\rightarrow\ket{e,0})$. The anharmonic coupling of the localized phonon modes to acoustic modes yields a decay channel with rate $\kappa_j^- = \kappa_j\left[N(\nu_j)+1\right]$ and an absorption channel with rate $\kappa_j^+ = \kappa_j N(\nu_j)$, where $N(\nu_j)=1/\left[\exp(\hbar\nu_j/k_B T)-1\right]$ denotes the thermal occupation of the phonon mode at temperature $T$.
Focusing on the regime of coherent emission, we assume low temperatures where $\hbar\nu_j \gg k_B T$. This allows us to neglect thermal phonon occupation ($\kappa^+_j\approx0$) and treat the phonon decay rate $\kappa_j^-\approx\kappa_j$ as constant . 
The remaining terms in the master equation [Eq.~\eqref{ME_pl}] correspond to spontaneous emission $\gamma$, and temperature dependent pure dephasing $\gamma_{\rm pd}$ driven by second-order electron-phonon interactions~\cite{Muljarov2004DephasingPhonons, Reigue2017ProbingDots, Tighineanu2018PhononDimensionality, Clear2020Phonon-InducedMolecules}.

To study the emission properties of the TLE, we calculate the first-order coherence function $\mathcal{G}^{(1)}(\tau) = \langle \sigma^\dagger(t)\sigma(t + \tau)\rangle_{\rm L}$ in the lab frame. Transforming to the polaron frame dresses the system transition operators with phonon displacement operators, yielding the modified two-time correlation function $\mathcal{G}^{(1)}(\tau) = \lim_{t\rightarrow\infty}\langle \sigma^\dagger(t)B_+(t)\sigma(t + \tau)B_-(t + \tau)\rangle_{\rm PF}$~\cite{Nazir2016ModellingDots, Iles-Smith2017PhononSources, Iles-Smith2017LimitsEmitters}. While this expression could be solved exactly by numerically integrating Eq.~\eqref{ME_pl} via the quantum regression theorem~\cite{Carmichael1999StatisticalEquations}, the approach scales exponentially with the number of phonon modes. 
However, accurate approximations can be obtained once we assume that phonon relaxation occurs much faster than spontaneous emission, which is typically valid for solid-state and molecular emitters~\cite{Clear2020Phonon-InducedMolecules, PhysRevResearch.6.033044}. This permits a mean-field approximation between phonon operators in the polaron frame and TLE operators, $ \mathcal{G}^{(1)}(\tau) \approx \langle B_+B_-(\tau)\rangle_{\rm PF}\lim_{t\rightarrow\infty} \langle \sigma^\dagger(t)\sigma(t + \tau)\rangle_{\rm PF} \equiv  \mathcal{G}_{\rm ph}(\tau) \mathcal{G}_{\rm pt}(\tau)$. This approximation enables an analytic solution which accurately reproduces the full numerical dynamics (see SI) while allowing the inclusion of an arbitrary number of mode.
Since the electronic transition operators remain approximately static on the fast $\sim 1/\kappa_j$ timescale, $\mathcal{G}_{\rm ph}(\tau)$ can be evaluated using only the phonon-relevant operators and dissipator terms of the master equation~\eqref{ME_pl}, with the form 

\begin{equation}
\mathcal{G}_{\rm ph}(\tau)= \mbox{exp}\left(-\sum_{j= 1}^M \beta_j\bigg(1 - e^{-(\kappa_j/2 + i\nu_j)\tau}\bigg)\right),
\label{corr_phon}
\end{equation}
where $\beta_i\equiv (\eta_i/\nu_i)^2$ is the Huang-Rhys factor. 
Combining this with the photonic part of the correlation function and neglecting the coherent scattering contribution, we obtain
\begin{equation}
\mathcal{G}^{(1)}(\tau) = \mathcal{G}_{\rm ph}(\tau) \left[ \frac{1}{4} e^{-\mathcal{C}_0\tau} + \sum_{\alpha = \pm1} \Lambda_\alpha e^{-\mathcal{S}^{(\alpha)}_{0}\tau} \right],
\label{eq:g1_gen}
\end{equation}
where $\mathcal{C}_0 = \Gamma/2 + i \tilde{\omega}_0$, $\mathcal{S}^{(\alpha)}_{0} = (3\gamma + \gamma_\mathrm{pd})/4 + i (\tilde{\omega}_0 + \alpha\tilde{\Omega}_\Gamma)$,
$\Lambda_\alpha = {\tilde{\Omega}^2}/({8\tilde{\Omega}_{\Gamma}^2 + 2\alpha i\tilde{\Omega}_\Gamma(\gamma - 2\gamma_\mathrm{pd})})$, $\tilde{\Omega} = \langle B_\pm\rangle \Omega$ is the phonon-renormalized Rabi frequency where $\langle B_{\pm} \rangle = \mbox{exp}(-\sum_{j= 1}^M \beta_j/2)$ is the Franck-Condon factor associated to the ZPL transition.
As in the atomic case~\cite{DelValle2011RegimesExcitation}, the frequency splitting of the Mollow triplet is given by $\tilde{\Omega}_{\Gamma} = (\tilde{\Omega}^2 - \left({\gamma}/{4} - {\gamma_{\rm pd}}/{2}\right)^2)^{1/2}$, a Mollow triplet is only observable when this quantity is real-valued, i.e.  $\tilde{\Omega} \geq |\gamma/4 - \gamma_{\rm pd}/2|$.

\emph{Single local phonon mode---}
To develop first insight, we initially restrict our discussion to a single localized phonon mode ($M=1$), a simplification that we will generalize later for complex molecules. Since the laser is resonant, higher phonon manifolds of excited state are not populated. Consequently, only the transitions of the form $\ket{e,0}\rightarrow\ket{g,n}$ are involved, implying a phonon-number change $\delta n \equiv n$.
Thus for this one phonon mode, the first-order coherence function in Eq.~\eqref{eq:g1_gen} can be written through a series expansion (see SI) as
\begin{equation}\mathcal{G}^{(1)}(\tau) = \sum_{n=0}^\infty \frac{e^{-\beta}\beta^n}{n!} \left[ \frac{1}{4} e^{-\mathcal{C}_n\tau} + \sum_{\alpha=\pm1} \Lambda_\alpha e^{-\mathcal{S}^{(\alpha)}_{n}\tau} \right],
\label{A_spectra}
\end{equation}
where $\mathcal{C}_n = \mathcal{C}_0 + n (\kappa/2 - i\nu)$ and $\mathcal{S}^{(\alpha)}_{n} = \mathcal{S}^{(\alpha)}_{0} + n (\kappa/2 - i\nu)$. Each term $n$ in this sum corresponds to a PSB replica, with its relative intensity governed by the Franck-Condon probability $e^{-\beta}\beta^n/n!$~\cite{Walker1979OpticalDiamond}. 
\begin{figure}
    \centering
    \includegraphics[width=0.9\linewidth]{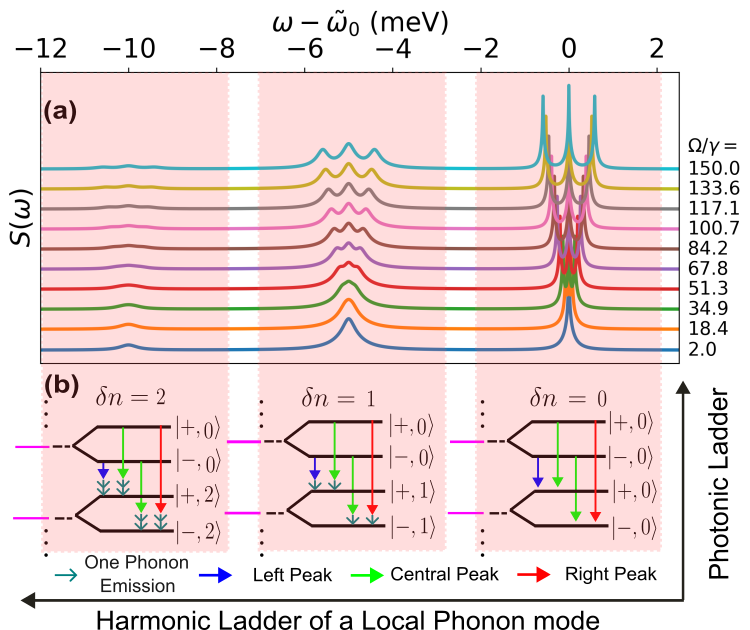}
    \caption{(a) Analytical spectra (normalized, log scale) of an emitter interacting with a single local phonon mode under varying driving strengths, revealing the emergence of vibronic Mollow triplets. The parameters used are $\nu = 5$~meV, $\eta = \nu/3$~meV and $\kappa = 0.2$~meV. (b) Transitions between dressed states explaining the appearance of Mollow triplets accompanied also by phonon generation resulting in not only the ZPL ($\delta n=0$), but also the first ($\delta n=1$) and second ($\delta n=2$) phonon sidebands. The diagram is not to scale, as the phonon energies exceed the Rabi splitting.}
    \label{fig:dressed}
\end{figure}
The first-order coherence function $\mathcal{G}^{(1)}(\tau)$ [Fig.~1(b)] reveals two distinct contributions: long-time oscillations at the renormalized Rabi frequency $\tilde{\Omega}$ yield the standard ZPL Mollow triplet [Fig.~1(c)]~\cite{Ramsay2010renorm,Wei2014renorm}, while short-time fluctuations arising from phonon relaxation generate the PSB replicas. Crucially, however, when the driving strength approaches the phonon relaxation rate, these fast phonon dynamics effectively modulate the electronic Rabi oscillations. This interplay is the physical origin of the new spectral features we report. The analytic structure of Eq.~\eqref{A_spectra} confirms this mechanism: each sideband term ($n\ge1$) mirrors the correlation function of the ZPL ($n=0$), differing only by the $n$-dependent terms $\mathcal{C}_n = \mathcal{C}_0 + n (\kappa/2 - i\nu)$ and $\mathcal{S}^{(\alpha)}_{n} = \mathcal{S}^{(\alpha)}_{0} + n (\kappa/2 - i\nu)$. Consequently, Eq.~\eqref{A_spectra} predicts a full \textit{vibronic Mollow triplet} imprinted on each PSB, comprising a central peak at $\tilde{\omega}_0 - n\nu$ with width $\Gamma_{\rm C} = \gamma +2\gamma_\mathrm{pd} + n \kappa$, and sidepeaks at $ \tilde{\omega}_0 - n\nu \pm \tilde{\Omega}_\Gamma$ with width $\Gamma_{\rm S} = (3\gamma + 2\gamma_\mathrm{pd} + 2n\kappa)/2$.
Fig.~\ref{fig:dressed} (a) demonstrates the emergence of this vibronic Mollow triplet within the PSBs as the Rabi frequency $\tilde{\Omega}$ increases, calculated using our analytical result in Eq.~\eqref{A_spectra}. These analytical spectra show excellent agreement with direct numerical evaluation of Eq.~\eqref{ME_pl} (see SI).
 
To elucidate origin of the vibronic Mollow triplets, we extend the standard dressed-atom picture~\cite{cohen1977dressed,cohen2024atom,Carmichael1999StatisticalEquations} to incorporate the vibrational manifold, as illustrated in Fig.~\ref{fig:dressed}~(b). This reveals a dual dressing of the emitter: the strong resonant field dresses the entire vibronic ladder, not merely the bare electronic ZPL. 
Consequently, transitions between these hybridized states allow photon emission to be accompanied by phonon generation, thereby replicating the Mollow triplet structure across the phonon sidebands.
Unlike previous proposals for tripartite electronic-vibrational-photonic states that rely on strong cavity coupling~\cite{Latini2021,Svendsen2023}, the vibronic dressed states reported here are generated dynamically by the incident driving field alone.

In contrast to the standard MT, observing the vibronic Mollow triplets requires satisfying a stricter resolvability condition due to the additional phonon decay rate $n\kappa$ given by $\tilde{\Omega}_\Gamma\geq (\Gamma_{\rm C} + \Gamma_{\rm S})/2$. Accounting for all decay channels, we derive the criterion for resolving the $n^{\rm th}$ vibronic triplet associated to the $j^\mathrm{th}$ phonon mode as:
\begin{equation}
\tilde{\Omega} \geq \tfrac{1}{4} \sqrt{(\gamma - 2\gamma_{\rm pd})^2 + (4 n_j \kappa_j + 5 \gamma + 6\gamma_{\rm pd})^2},
\label{rabi_bound}
\end{equation}
which we place in a practical experimental context in the subsequent section on DBT.

Despite sharing a common origin, vibronic Mollow triplets in fact exhibit distinct spectral characteristics compared to their ZPL counterpart. We quantify this via the linewidth ratio $R_{\Gamma}=\Gamma_{\rm C}/\Gamma_{\rm S}$ and amplitude ratio $R_{\rm A} = A_{\rm C}/A_{\rm S}$ as depicted in the inset of Fig.~\ref{fig:main}(c). From Eq.~\eqref{A_spectra}, generalized for $j$-th local phonon mode ($n\kappa\rightarrow n_j\kappa_j$), we obtain:
\begin{equation}
R_{\Gamma_j} = \frac{\gamma + 2\gamma_{\rm pd} + n_j \kappa_j}{1.5\gamma + \gamma_{\rm pd} + n_j\kappa_j}, \quad R_{\mbox{A}_j} \approx \left(\frac{2}{R_\Gamma}\right)\left|\frac{1}{8\Lambda_{\alpha}}\right|.
\label{eq:Rgamma}
\end{equation}
In the bare atomic limit ($\eta=0$), we recover the canonical values $R_\Gamma = 1:1.5$ (sidepeaks are broader) and $R_{\rm A} = 3:1$ (sidepeaks are lower)~\cite{Carmichael1999StatisticalEquations}, as shown in Fig.~\ref{fig:main}(c).  In contrast, the vibronic triplets are dominated by the phonon decay $n_j\kappa_j$, which effectively equalizes the linewidths ($R_\Gamma \to 1:1$) and modifies the amplitude ratio to $R_{\rm A} \approx 2:1$ (assuming $|8\Lambda_\alpha|\approx 1$). This relative narrowing of the sidepeaks [Fig.~\ref{fig:dressed}(a)] is a direct signature of the additional phonon relaxation pathway, which acts as a common broadening mechanism for all vibronic triplets.

\emph{Emission spectrum of DBT---}
We now extend our analysis to multiple phonon modes, allowing us to model realistic molecular emitters like DBT. DBT is a promising single-photon source candidate, distinguished by a strong, near-transform-limited ZPL~\cite{Nicolet, Zirkelbach2022High-resolutionCrystal}. Its electronic states couple to ten primary localized vibrational modes, well-characterized by optical~\cite{Clear2020Phonon-InducedMolecules, Zirkelbach2022High-resolutionCrystal} and Raman~\cite{PhysRevLett.127.123603} spectroscopy. Crucially, the observation of the conventional ZPL Mollow triplet in DBT~\cite{wrigge2008efficient} confirms that the requisite coherent driving regime is experimentally accessible, making it an ideal platform to search for vibronic Mollow triplets.

\begin{figure}
    \centering
    \includegraphics[width=\linewidth]{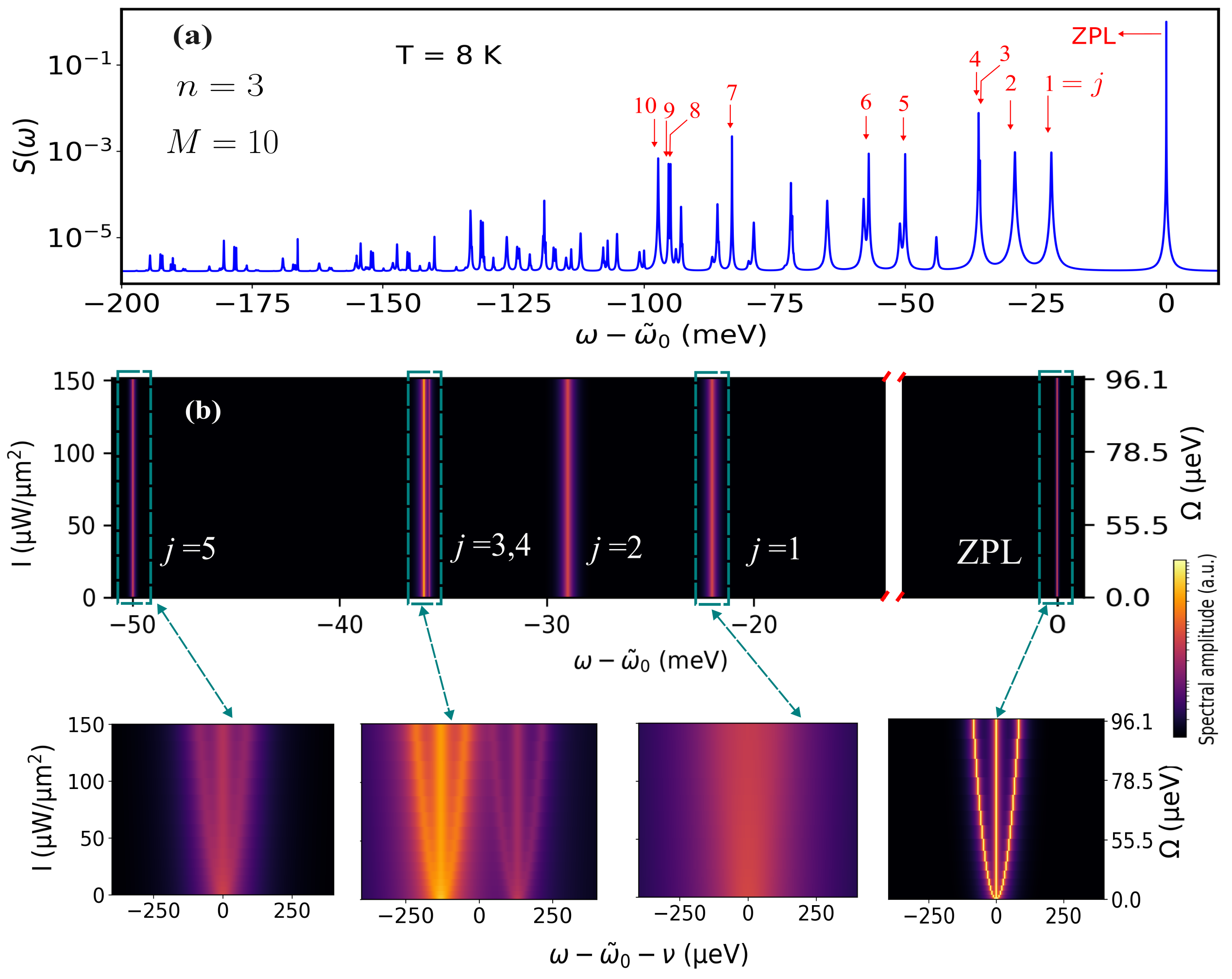}
    \caption{(a) Normalized DBT spectrum from \eqref{eq:g1_gen} at $T=8~\mathrm{K}$ shows ten fundamental modes (blue; red arrows mark peak positions by mode index) their higher-order overtones up to $n=3$ and sums of mode frequencies (in pink). Other simulation parameters are $\gamma = 0.094\;\text{µ}$eV and $\Omega = 10\gamma$. (b) Laser intensity (left $y$-axis) and its corresponding Rabi frequency (right $y$-axis) dependent lifetime limited DBT spectra for the first five fundamental local phonon modes, computed from \eqref{eq:DBT_ana}. The bottom panel presents zoomed-in views of modes $j=1,3,4,5$. We use $\lambda = 745$ nm and the transition dipole moment as $d = 13.7D$ where $D =3.34\times 10^{-30}$ C-m~\cite{Zirkelbach2022High-resolutionCrystal}.}
    \label{fig:DBT_spectra}
\end{figure}

Generalizing to a multi-mode system transforms the emission spectrum into summation over all modes $j$ and their corresponding phonon replicas $n_j$. 
Adopting experimental values from Zirkelbach et al.~\cite{Zirkelbach2022High-resolutionCrystal}, we explicitly model ten phonon modes characterized via Raman spectroscopy of DBT in para-dichlorobenzene (see SI Table II for parameters).
The resonance fluorescence for DBT driven at $\Omega =10\gamma$ is given in Fig.~\ref{fig:DBT_spectra}(a) demonstrating the dominant ZPL, and ten local phonon modes ($M=10$) and their three associated phonon replicas ($n=3$), as well as resonances corresponding to combinations of frequency sums.
A significant advantage of our analytical formalism, however, is that it allows for selecting only the dominant emission frequencies of the $n_j \in \{0, 1\}$ manifold. This captures the full dynamics of the ZPL and the first PSB of each mode without further approximation. This capability provides a crucial benefit over standard numerical master equation treatments, where truncating the vibronic Hilbert space—a step required for computational tractability—often introduces errors unless a sufficient number of states are accounted for. 
Our method thus provides an accurate and scalable framework for analyzing the experimentally dominant ZPL and first-order PSBs in complex molecular systems.
With this assumption in Eq.~\eqref{corr_phon} and following a similar procedure for evaluating the photonic part of the correlation function we obtain,
 \begin{equation}
     \mathcal{G}^{(1)}(\tau) = \mbox{e}^{-\tilde{\beta}} \sum_{j= 0}^M\beta_j  \Bigg[ \frac{1}{4} e^{-\mathcal{C}_j\tau} + \sum_{\alpha = \pm1}\Lambda_\alpha e^{-\mathcal{S}^{(\alpha)}_{j}\tau}\Bigg],
     \label{eq:DBT_ana}
 \end{equation}
where we defined $\beta_0 \equiv 1$ to write the expression in compact form. Also here $\tilde{\beta} = \sum_{j= 1}^M \beta_j$,   $\mathcal{C}_j = \mathcal{C}_0 + (\kappa_j/2  - i\nu_j)$,  $\mathcal{S}^{(\alpha)}_{j} = \mathcal{S}^{(\alpha)}_{0} + (\kappa_j/2 - i\nu_j)$.  The case, $j = 0$ corresponds to the ZPL, whose spectral weight is reduced by a factor of $\mbox{exp}(-\tilde{\beta})$ relative to the bare atomic case representing also the dechoherence due to the local phonons. In contrast, $j>0$ represent phonon-assisted transitions, leading to sideband peaks at frequencies $\tilde{\omega}_0 - \nu_j$, with their respective amplitudes suppressed by a factor of $\beta_j\mbox{exp}(-\tilde{\beta})$. 
The Rabi frequency in this case is renormalized by the contributions of all the relevant modes thus $\tilde{\Omega} =\Omega\;\mbox{exp}(-\tilde{\beta}/2)$.  

We assess experimental feasibility in Fig.~\ref{fig:DBT_spectra}(b), which maps the vibronic splitting of five DBT modes [Eq.~\eqref{eq:DBT_ana}] to laser intensity (left axis) and equivalent Rabi frequency (right axis; see SI for calibration). The top panel details individual modes: the strongly damped modes $j=1,2$ ($\kappa_{1,2} \approx 130\text{--}160~\text{µeV}$) exceed the maximum simulated drive, preventing triplet resolution. Conversely, the lower-damping modes $j=3\text{--}5$ ($\kappa \approx 35\text{--}55~\text{µeV}$) exhibit distinct splitting, consistent with the lifetime-limited bound $\tilde{\Omega} \gtrsim n_j\kappa_j$ [Eq.~\eqref{rabi_bound}].
  The drive strengths (Rabi frequencies) required to resolve the vibronic Mollow splitting in our example are comparable to those reported in contemporary demonstrations of Mollow triplets~\cite{Unsleber2015ObservationDot, gerard2025res}, though in other platforms. In particular, a drive strength of $\tilde{\Omega}\sim 35~\text{µeV}$ is sufficient to observe the vibronic triplet corresponding to laser intensity of $20~\text{µW}/\text{µm}^{2}$. Nevertheless, very high optical powers can induce photobleaching \cite{Christ2001}, charge-state fluctuations \cite{Duquennoy2024} and spectral dephasing~\cite{Konthasinghe:19, Wang2022}. Materials engineering can mitigate these effects for instance, using hexagonal boron nitride~\cite{Han2021, smit2023}, electrical gating \cite{Akbari2022Lifetime-LimitedModulation, Paralikis2025} which improves environmental stability and reduces charge noise, thereby enabling higher drive without degradation. Moreover, prolonging the lifetimes of phonon modes~\cite{PhysRevLett.127.123603} would lower the threshold drive required to resolve vibronic Mollow triplets—consistent with the condition $\tilde{\Omega}\gtrsim \kappa_j$ for lifetime limited emitters, thereby improving their experimental feasibility.

\textit{Conclusion---} In conclusion, we identify the emergence of \textit{vibronic Mollow triplets}: Mollow triplets in the local phonon sidebands of a strongly driven quantum emitter. These features signal the formation of tripartite dressed states, generated by the coherent interplay between the emitter, the laser, and the vibrational environment.  By extending the canonical dressed-atom framework, we derive analytical expressions that exhibit excellent agreement with numerical polaron master equations while highlighting the distinct spectral constraints of the vibronic regime given by Eq.~\eqref{rabi_bound}. This formalism overcomes the scaling limitations of numerical methods, enabling the simulation of complex systems like DBT. Crucially, while demonstrated here for DBT, our findings apply to any quantum emitter coupled to high-$Q$ localized phonon modes. Ultimately, our results highlight that the phonon sideband is not only an incoherent loss channel, but under appropriate driving conditions may be considered coherent spectral resource, 
opening new avenues for quantum control across diverse solid-state platforms.

\emph{Acknowledgments---} We would like to thank Alex Clark and Moritz Cygorek for stimulating conversations. This work was supported by the Danish National Research Foundation through NanoPhoton—Center for Nanophotonics, Grant No. DNRF147. D.P., C.K., and N.S. acknowledges funding by the Novo Nordisk Foundation NERD Programme (Project QuDec NNF23OC0082957). 

\bibliography{refs_clean}
\bibliographystyle{unsrt}

\clearpage
\onecolumngrid
\appendix

\pagebreak
\widetext
\begin{center}
\textbf{\large Supplementary Information: Resonant Excitation Induced Vibronic Mollow Triplets}
\end{center}

\setcounter{figure}{0}
\setcounter{equation}{0}
\setcounter{table}{0}
\setcounter{page}{1}
\makeatletter
\renewcommand{\theequation}{S\arabic{equation}}
\renewcommand{\thefigure}{S\arabic{figure}}
\renewcommand{\bibnumfmt}[1]{[S#1]}
\renewcommand{\citenumfont}[1]{S#1}

\title{Resonant Excitation Induced Vibronic Mollow Triplets\\ (Supplemental Information)} %
%
            
\maketitle            


\maketitle
\section{Derivation of Polaron master equation}
In this section we derive a master equation used in the main text. In the frame rotating with respect to the laser we have,
\begin{equation}
    H' = H_{\rm e}' + H_{\rm ph} + H_{\rm e-ph} + H_{\rm pt} + H_{\rm e-pt}' + H'_{\rm laser}
\end{equation}
where $ H_{\rm e}' = \delta \sigma^\dagger\sigma$, where $\delta = \omega_0 - \omega_l$ , $H_{\rm e-pt}' =  \sum_q\big(h_q a_{q}^\dagger\sigma e^{-i\omega_l t} + h_q^*  a_{q}\sigma^\dagger e^{i\omega_l t}\big)$ and $H'_{\rm laser} = \Omega\sigma_{x}/2$. We also define the total state as a composite state $\chi(0) = \rho_s(0)\otimes\rho_E$ where the system state comprises as the state of the emitter and the phonons given by $\rho_s = \rho_a\otimes\rho_{\rm \nu}$ and the electromagnetic environment state is given by a time independent state $\rho_E$.
Taking polaron transform  of the above Hamiltonian as $\tilde{H}' = U_{Pol}H'U_{Pol}^\dagger$, through a unitary transformation $U_{Pol} =|g\rangle\langle g| + |e\rangle\langle e|B_+$ where $B_+ = \mbox{exp}(\sum_{j = 1}^M\eta_j^2/\nu_j^2(b_j^\dagger - b_j))$ we obtain,
\begin{equation}
 \tilde{H}' =   \underbrace{\tilde{\delta} \sigma^\dagger\sigma + \sum_{j= 1}^M\nu_j b_j^\dagger b_j +  \Omega(\sigma B_- + \sigma^\dagger B_+)/2 + \sum_{q}\omega_{q} a_{q}^\dagger a_{q}}_{\tilde{H}'_0 = \tilde{H}_s + H_{\rm pt}}  \\ + \underbrace{\sum_q\big(h_q a_{q}^\dagger \sigma_p e^{-i\omega_l t} + h_q^*  a_{q}\sigma_p^\dagger e^{i\omega_l t}\big)}_{\tilde{H}_I'} 
\end{equation}
We have also segregated the free Hamiltonian $\tilde{H}_0$ into the system Hamiltonian $\tilde{H}_s$ and the photon environment Hamiltonian $H_{\rm pt}$. The Born-Markov master equation in the interaction picture is thus given by,
 \begin{equation}
   \frac{d\rho_I(t)}{dt} = -\int_0^\infty d\tau \mbox{Tr}_E \bigg(\big[\tilde{H}'_I(t), \big[\tilde{H}'_I(t-\tau),\rho_I(t)\otimes \rho_E\big]\bigg)  
 \end{equation}
 Therefore substituting the interaction Hamiltonian in the interaction picture as $\tilde{H}_{I}'(\tau) = U_0^\dagger(\tau) \tilde{H}_I' U_0(\tau) = \sum_{i= 1}^2 O_{si}(\tau)\otimes E_i(\tau)$, where $ O_{si}(\tau) = e^{i \tilde{H}_s \tau} O_{si} e^{-i \tilde{H}_s \tau} $ with $O_{s1} = O_{s2}^\dagger$ and $E_1(\tau) = \sum_q h_q a_q^\dagger e^{i(\omega_q - \omega_l)\tau} = E_2^\dagger(\tau)$, in the above master equation and then transforming into the Schrödinger picture we obtain,

 \begin{equation}
     \frac{d\rho_s(t)}{dt} = -i[\tilde{H}_s, \rho_s(t)] -\sum_{i \neq j \in (1,2)}\int_0^\infty d\tau \bigg(\big[O_{si}, O_{sj}(-\tau)\rho_s(t) \big] \mathcal{C}_{ij}(\tau) + \big[\rho_s(t)O_{sj}(-\tau), O_{si}\big]\mathcal{C}_{ji}(-\tau)\bigg)
\end{equation}
 where $\mathcal{C}_{ij}(\tau) = \mbox{Tr}_E\big[E_i E_j(-\tau) \rho_E\big]$ and we find that $\mathcal{C}_{11} = \mathcal{C}_{22} = \mathcal{C}_{12} = 0$ in the case of the environment state as the multimode vacuum state while $\mathcal{C}_{21}(\tau) = \sum_q |h_q|^2 e^{-i(\omega_q - \omega_l)\tau} = \int_0^\infty J_{\rm pt}(\omega)e^{-i(\omega - \omega_l)\tau}$ where $J_{\rm pt}(\omega) = \sum_q|h_q|^2\delta(\omega - \omega_q)$ is the photon spectral density. The above equation can be written in a compact form as follows.
 \begin{equation}
     \frac{d\rho_s(t)}{dt} = -i[\tilde{H}_s, \rho_s(t)] -\left( \big[O_{s2}, \beta_{1}\rho_s(t)\big] + \big[\rho_s(t)\beta_{2}, O_{s1}\big]\right)
 \end{equation}
 where $\beta_{1} = \int_0^\infty d\tau O_{s1}(-\tau)\mathcal{C}_{21}(\tau)$ and $\beta_{2} = \int_0^\infty d\tau O_{s2}(-\tau)\mathcal{C}_{21}(-\tau)$. Now using the flat spectral density approximation i.e. $J_{\rm pt}(\omega) \approx \zeta$, the $\beta_1$ and $\beta_2$ coefficients can be given as follows,
 \begin{equation}
     \beta_1 = \zeta\pi O_{s1} - i\mathcal{P}\left(\zeta\int_0^\infty d\tau\frac{ O_{s1}(-\tau)e^{i\omega_l\tau}}{\tau}\right),\;\beta_2 = \zeta\pi O_{s2} + i\mathcal{P}\left(\zeta\int_0^\infty d\tau\frac{ O_{s2}(-\tau)e^{-i\omega_l\tau}}{\tau}\right)
 \end{equation}
 We thus obtain the polaron master equation as follows,

 \begin{eqnarray}
     \frac{d\rho_s(t)}{dt} &=& -i[\tilde{H}_s, \rho_s(t)] + \gamma\left(O_{s1}\rho_s(t)O_{s2} - \frac{1}{2}\{O_{s2}O_{s1}, \rho_s(t)\}_+\right)\\
     &=& -i[\tilde{H}_s, \rho_s(t)] + \gamma\left(\sigma B_-\rho_s(t)\sigma^\dagger B_+ - \frac{1}{2}\{\sigma^\dagger\sigma, \rho_s(t)\}_+\right)
 \end{eqnarray}
 where $\tilde{H}_s = \tilde{\delta} \sigma^\dagger\sigma + \sum_{j= 1}^M\nu_j b_j^\dagger b_j +  \Omega(\sigma B_- + \sigma^\dagger B_+)/2$ and $\tilde{\delta} = \omega_0 - \omega_l - \sum_{j = 1}^M\eta_j^2/\nu_j$. Adding also localized phonon decay (decaying anharmonically to acoustic phonons) with the decay rate given by $\kappa_j(N(\nu_j) + 1)$ as well as the phonon absorption with the rate $\kappa_jN(\nu_j)$ where $N(\nu_j) = 1/(\mbox{exp}(\hbar\nu_j/k_BT - 1))$ is the occupation number for the phonons and $\kappa_j \propto (\nu_j^3/\omega_c^2) \mbox{exp}(-\nu_j/\omega_c)$ where $\omega_c$ is the acoustic phonon environment cutoff frequency \cite{Groll2021ControllingProposal, Clear2020Phonon}. Furthermore, we note that in the temperature regime of interest, the phonon occupation number is negligible as shown in Fig.~\ref{fig:DBT_spectra_dephasing}(a), allowing us to drop the  the phonon absorption rate as well as treat the phonon decay rate as independent of temperature. Finally we also add the dephasing rate $\gamma_{\rm pd}$ phenomenologically that encapsulates the effect of dephasing due to the acoustic phonon environment. We thus obtain the following master equation,
 \begin{multline}
     \frac{d\rho_s(t)}{dt} = -i[\tilde{H}_s, \rho_s(t)] + \gamma\left(\sigma B_-\rho_s(t)\sigma^\dagger B_+ - \frac{1}{2}\{\sigma^\dagger\sigma, \rho_s(t)\}_+\right) + \sum_{j=1}^M\kappa_j\left(b_j\rho_s(t)b^\dagger_j - \frac{1}{2}\{b^\dagger_j b_j, \rho_s(t)\}_+\right)\\ + 2\gamma_{\rm pd}\left(\sigma^\dagger\sigma\rho_s(t)\sigma^\dagger\sigma - \frac{1}{2}\{\sigma^\dagger\sigma, \rho_s(t)\}_+\right) \\ 
     \label{ME_pl_app}
 \end{multline}
 \section{Equations of motion of expectation values and steady state values}
 From the above equation we can derive the steady state values of the operators that follow the following equations of motion~\cite{steck2007quantum},
\begin{equation}
    \frac{d\langle \sigma \rangle}{d\tau} = -\frac{\Gamma}{2}\langle \sigma \rangle + \frac{i\tilde{\Omega}}{2}\langle \sigma_z \rangle,\;\;\frac{d\langle \sigma^\dagger\rangle}{d\tau} = -\frac{\Gamma}{2}\langle \sigma^\dagger \rangle - \frac{i\tilde{\Omega}}{2}\langle \sigma_z \rangle,\;\; \frac{d\langle \sigma_z\rangle}{d\tau} = i\tilde{\Omega}\langle \sigma \rangle - i\tilde{\Omega}\langle \sigma^\dagger\rangle - \gamma\langle \sigma_z\rangle - \gamma
    \label{sm_app}
\end{equation}
 where $\Gamma \equiv \gamma + 2\gamma_{\rm pd}$ and $\tilde{\Omega} = \langle B_{\pm} \rangle \Omega$ is the phonon renormlaized Rabi frequency due to the local phonon modes. The steady state value can be given as follows,
 \begin{equation}
     \langle \sigma^\dagger\sigma\rangle_{ss} = \rho_{ee}(\tau\rightarrow \infty) = \frac{1}{2}\frac{s}{s + 1},\;\;\langle \sigma_z\rangle_{ss} = -\frac{s}{s + 1},\;\;\langle \sigma^\dagger\rangle_{ss} = \rho_{ge}(\tau\rightarrow \infty) = i\sqrt{\frac{a s}{2}}\frac{1}{s + 1}
 \end{equation}
 where we define, $s = {2\tilde{\Omega}^2}/{\gamma\Gamma},\; a = {\gamma}/{\Gamma}$
 
 \section{Deriving the spectrum}
In the assumption that the phonon dynamics is much faster than the atomic (this assumption will not apply at very high driving and simultaneous large electron-phonon coupling) the correlation function in the polaron frame (PF) is evaluated as,
 \begin{equation}
     \langle \sigma^\dagger_p(t)\sigma_p(t + \tau)\rangle_{\rm PF} \approx \langle B_+(t) B_-(t + \tau)\rangle_{\rm PF}\langle \sigma^\dagger(t)\sigma(t + \tau)\rangle_{\rm PF}
 \end{equation}
 \subsection{Evaluating the phononic part of the correlation}
 
 We can evaluate the phonon correlation using the master equation \eqref{ME_pl_app} by considering only the phonon relevant operators and dissipator which we can rewrite in the form of a non-Hermitian effective Hamiltonian given by $H_{eff} = \sum_{j= 1}^M(\nu_j - i\kappa_j/2)b_j^\dagger b_j$. In the assumption of $N(\nu)\rightarrow0$ the thermal state can be approximated to the vacuum phonon state $\rho_{\nu}^{ss} = | {0}\rangle \langle {0}| $ ($B_-| {0}\rangle \langle {0}| B_+$ in the lab frame) gives us the following form of phonon correlation,
 \begin{eqnarray}
 \langle B_+(t) B_-(t + \tau)\rangle_{\rm PF}|_{t\rightarrow \infty}&=&\prod_{i=1}^M\langle B^{(i)}_+(t) B^{(i)}_-(t + \tau)\rangle\equiv \mathcal{G}_{\rm ph }(\tau)\\ &=& \mbox{exp}\left(-\sum_{j= 1}^M \beta_j\bigg(1 - e^{-(\kappa_j + i\nu_j)t}\bigg)\right)
 \label{corr_phon_app}
 \end{eqnarray}
where $\beta_i \equiv |\eta_i/\nu_i|^2$. 
 \subsection{Evaluating the atomic part of the correlation}
 The atomic correlation can be written as the sum of coherent and incoherent scattering part as $\langle\sigma^\dagger(t)\sigma(t + \tau)\rangle = \langle\delta\sigma^\dagger(t)\delta\sigma(t + \tau)\rangle + \langle\sigma^\dagger\rangle_{ss}\langle\sigma\rangle_{ss}$ where $\delta\sigma \equiv \sigma - \langle\sigma\rangle_{ss}$. The coherent part can be evaluated as,
 \begin{equation}
   \langle\sigma^\dagger\rangle_{ss}\langle\sigma\rangle_{ss} = |\rho_{eg}(t\rightarrow\infty)|^2 = \frac{1}{2}\frac{a s}{(s + 1)^2 }  
   \label{corr_coh_app}
 \end{equation}
 From quantum regression theorem $\langle\sigma^\dagger(t)\sigma(t + \tau)\rangle_{t\rightarrow \infty} = \rm{Tr_s}[\sigma \Lambda(\tau)] =  \langle e |\Lambda(\tau)| g\rangle \equiv \Lambda_{eg}(\tau)$, where $\Lambda(\tau) = \rm{Tr_E}[U(\tau)\chi_{ss}\sigma^\dagger U^\dagger(\tau)]$ where $U(\tau)$ is the unitary evolution and $\chi_{ss}\equiv \chi(t\rightarrow \infty)$ being the composite steady state of the system and environment. $\Lambda(0) = \rho_s(t\rightarrow \infty)\sigma^\dagger$ (which is subjected to the assumption that at the steady state the states of the system and environment can be written in a separable form i.e. $\chi_{ss} = \rho_s(t\rightarrow\infty)\otimes \rho_E$). To evaluate the incoherent spectrum we essentially need an evolution equation of $\delta\Lambda(\tau)$ which corresponds to the state after we subtract the steady state components. From quantum regression theorem the density matrix $\delta\Lambda(\tau)$ follows the same evolution equation as the system density matrix i.e. $d\delta\Lambda(\tau)/d\tau = \mathcal{L}\delta\Lambda(\tau)$ with the initial state $\delta\Lambda(0) = \rho_s(t\rightarrow \infty)\delta\sigma^\dagger$. The Lindbladian $\mathcal{L}$, was already calculated in the previous section but here we see a slight difference with respect to the absence of the constant part of the matrix equation since we subtract the steady state values,
 \begin{equation}
 \mathcal{L} = 
 \begin{bmatrix}
\frac{-\Gamma}{2} & 0 & \frac{i\tilde{\Omega}}{2}\\
0 & \frac{-\Gamma}{2} & -\frac{i\tilde{\Omega}}{2}\\
i\tilde{\Omega} & -i\tilde{\Omega} & -\gamma
\end{bmatrix}
\end{equation}
 Therefore we will need to evaluate $\delta\Lambda(\tau) = \mbox{exp}(\mathcal{L}\tau)\delta\Lambda(0)$, particularly $\delta\Lambda_{eg}(\tau)$ where $\delta\Lambda_{\alpha\beta}(0) =\delta_{g\beta}\rho_{\alpha e}(t\rightarrow\infty) - \rho_{\alpha \beta}(t\rightarrow\infty) \rho_{g e}(t\rightarrow\infty)$~\cite{steck2007quantum} which gives us,
 \begin{equation}
 \delta\Lambda(0) = \frac{s}{2(s + 1)^2} 
     \begin{bmatrix}
s + 1 - a \\
a\\
-ia\sqrt{2s}
\end{bmatrix} \approx \begin{bmatrix}
{1}/{2} \\
{a}/{2s}\\
{-ia}/{\sqrt{2s}}
\end{bmatrix}, \;\;\; s\gg 1
 \end{equation}
Since $a\in (0, 1)$ and $s\gg1$ we can further approximate $\delta\Lambda(0) \approx \big(\begin{smallmatrix}
  1/2\\
  0\\ 0
\end{smallmatrix}\big)$ which gives us,
 \begin{equation}
     \langle\delta\sigma^\dagger(t)\delta\sigma(t + \tau)\rangle_{\rm PF}|_{t\rightarrow\infty} \equiv \mathcal{G}_{Pt}(\tau) =  \frac{1}{4}e^{-\Gamma \tau/2} +\sum_{\alpha\in\{-1,1\}} \Lambda_\alpha e^{-(2\gamma + \Gamma)\tau/4}e^{\alpha i\tilde{\Omega}_{\Gamma}\tau}
     \label{corr_atom_app}
 \end{equation}
 where 
 \begin{equation}
  \tilde{\Omega}_{\Gamma} = \sqrt{\tilde{\Omega}^2 - \left(\frac{\gamma}{4} - \frac{\gamma_{\rm pd}}{2}\right)^2},\; \text{and}\; \Lambda_\alpha = {\tilde{\Omega}^2}/({8\tilde{\Omega}_{\Gamma}^2 + 2\alpha i\tilde{\Omega}_\Gamma(2\gamma - \Gamma)})  
  \label{eq: Mollow_params}
 \end{equation}
  
 So for maximum Mollow splitting $\gamma_{\rm pd} = \gamma/2$ and not zero. The total correlation function can now be written as
 $\mathcal{G}^{(1)}(\tau) = \mathcal{G}_{\rm ph}(\tau) \mathcal{G}_{\rm pt}(\tau)$. Therefore, using Eqs.~\eqref{corr_phon_app}, \eqref{corr_coh_app} and \eqref{corr_atom_app} we get the combined correlation function as follows,
 \begin{equation}
  \mathcal{G}^{(1)}(\tau)_{t\rightarrow\infty} = \mbox{exp}\left(-\sum_{j= 1}^M \beta_j\bigg(1 - e^{-(\kappa_j + i\nu_j)t}\bigg)\right)\Bigg(\frac{a s}{2(s + 1)^2 } + \frac{1}{4}e^{-\Gamma \tau/2} + \sum_{\alpha\in \{-1,1\}}\Lambda_\alpha e^{-(2\gamma + \Gamma)\tau/4}e^{\alpha i\tilde{\Omega}_{\Gamma}\tau}\bigg)
  \label{A_spectra_app}\textbf{}
 \end{equation}
 In the paper we have discussed two cases and thus we will reformulate Eq.~\eqref{A_spectra_app} accordingly.\newline
 \textbf{Case I} : We consider only one mode i.e. $M = 1$. In this case the total correlation function leaving out the coherent part becomes,
 \begin{equation}
  \mathcal{G}^{(1)}(\tau)|_{t\rightarrow\infty} = \sum_{n = 0}^\infty \frac{\mbox{e}^{-\beta}\beta^n}{n!}\Bigg(\frac{1}{4}e^{-(\Gamma_{\rm C} + i\Omega_{\rm C}) \tau} + \sum_{\alpha \in \{-1,1\}} \Lambda_\alpha e^{-(\Gamma_{\rm S} + \alpha i\Omega_{\rm S})\tau}\Bigg) 
  \label{A_spectra_1mode}
 \end{equation}
 where $\Gamma_{\rm C} = (\Gamma + n \kappa)/2$, $\Gamma_{\rm S} = (2\gamma + \Gamma +2n\kappa)/4$. The term $\mbox{exp}(-\beta)\beta^n/n!$ can be recognized as the transition probability of the $n$-th vibronic level of the ground state of the single phonon mode given by $|\langle 0| B_\pm|n\rangle|^2$. $n = 0, 1 , 2...$, gives respectively the zero phonon line, the first phonon sideband and the second phonon sideband and so on. Thus for a single mode case as here, $n$ tells us about an infinite number of vibronic levels. In numerical simulation therefore we truncate the range of $n$ to a reasonable limit after which the contribution from the sidebands becomes negligible. In the case of simulation pertaining to the defects in hBN for instance we truncate the number of levels to $n = 3$. 
 \textbf{Case II} : We consider only one phonon replica i.e. $n \in(0, 1)$ and $M$ modes. In this case the total correlation function becomes,
 \begin{multline}
     \mathcal{G}^{(1)}(\tau)|_{t\rightarrow\infty} = \mbox{e}^{-\tilde{\beta}} \Bigg[ \frac{1}{4}\Bigg(e^{-(\Gamma_{{\rm C}_0} + i \Omega_{{\rm C}_0})\tau} + \sum_{j=1}^M \beta_j(e^{-(\Gamma_{{\rm C}_j} + i \Omega_{{\rm C}_j})\tau}\Bigg)\\ + \sum_{\alpha \in \{-1,1\}}\Lambda_\alpha\Bigg(e^{-(\Gamma_{{\rm S}_0} + \alpha i\Omega_{{\rm S}_0})\tau} + \sum_{j=1}^M \beta_j e^{-(\Gamma_{{\rm S}_j} + \alpha i\Omega_{{\rm S}_j})\tau}\Bigg)\Bigg]
 \end{multline}
 where $\tilde{\beta} = \sum_{j= 1}^M \beta_j$, $\Gamma_{{\rm C}_j} = (\Gamma + \kappa_j)/2$,  $\Omega_{{\rm C}_j} = \tilde{\omega}_0 - \nu_j$, $\Gamma_{{\rm S}_j} = (2\gamma + \Gamma + 4\kappa_j )/4$ and $\mp\Omega_{{\rm S}_j} = \tilde{\omega}_0 - \nu_j \pm \Omega_\Gamma$. Finally, we define the first order correlation function as:
 \begin{equation}
     g^{(1)}(\tau) = \frac{ \langle \sigma^\dagger_p(t)\sigma_p(t + \tau)\rangle_{t\rightarrow \infty}}{ \langle \sigma^\dagger_p(t)\sigma_p(t )\rangle_{t\rightarrow \infty}}
 \end{equation}
 where in the strong driving case the denominator $\langle \sigma^\dagger_p(t)\sigma_p(t )\rangle|_{t\rightarrow \infty} = \langle\sigma^\dagger\sigma\rangle_{ss} = 1/2$. Furthermore, from the Wiener–Khinchin theorem the spectrum is obtained as $S(\omega) = \int_{-\infty}^{\infty} d\tau g^{(1)}(\tau) \mbox{exp}(i\omega\tau)$.
\section{Temperature dependent analysis}
\begin{figure}[H]
    \centering
    \includegraphics[width=0.6\linewidth]{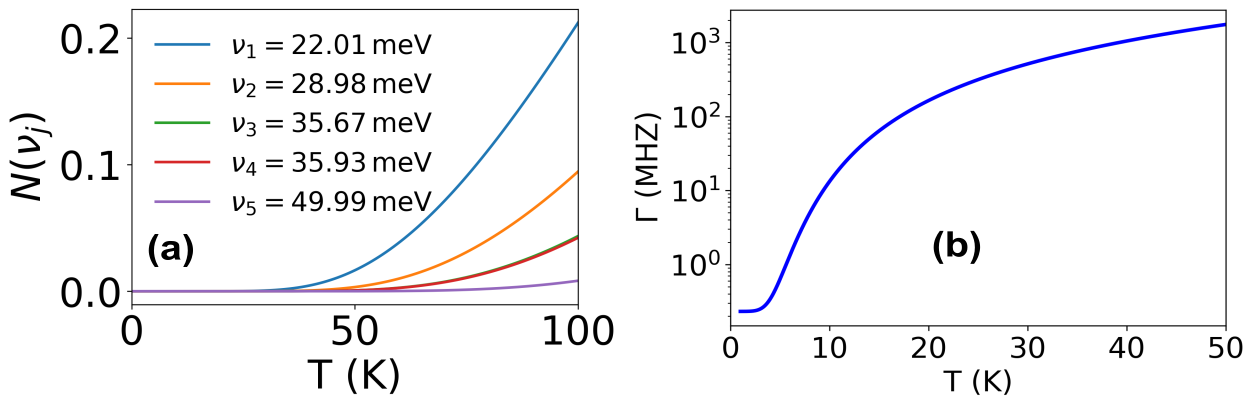}
    \caption{(a) Temperature regime showing upto where one can use $N(\nu_j)\approx0$ for different DBT modes in \textit{para}-dichlorobenzene given in Table.~\ref{tab:vibrational_modes_pCB}. (b) $\Gamma$ vs temperature extracted from the analytical result of \cite{Clear2020Phonon}. }
    \label{fig:DBT_spectra_dephasing}
\end{figure}
The semi-analytic expression employed in Fig.~\ref{fig:DBT_spectra_dephasing}(b) and used to model the temperature-dependent dephasing rate $\gamma_{\rm pd}$ in the main text is given by,
\begin{equation}
\gamma_{\rm pd} = \mu\int d\omega \omega^6 N(\omega)(N(\omega) +1)\mathcal{K}(\omega),\;\text{where},\; \mathcal{K}(\omega) = \frac{8\left(3 - e^{-2 a(\omega)} \left[ 3 + 6 a(\omega) + 6 a(\omega)^2 + 4 a(\omega)^3 + 2 a(\omega)^4 \right]\right)}{a(\omega)^5}    
\label{eq: dephasing}
\end{equation}
with $a(\omega) = 2(\omega/\omega_c)^2$. The numerical values of the constants used in the simulations are $\omega_c = 8.6\; \text{ps}^{-1} (5.67\;  \text{meV})$ and $\mu = 4.7\times10^{-7}\;\text{ps}^5$.
\begin{figure}[h!]
    \centering
    \includegraphics[width=0.7\linewidth]{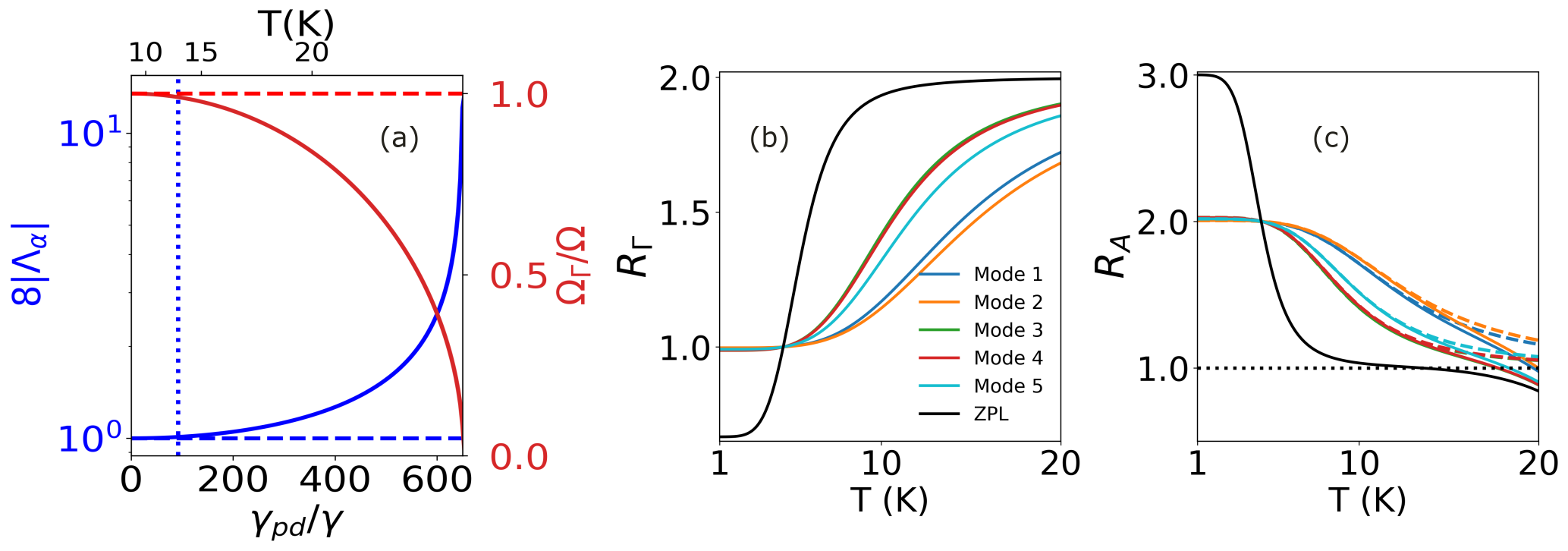}
    \caption{(a) Regime of validity the $8\Lambda_\alpha= 1$ approximation for $\Omega = 2\kappa_2$ (Table \ref{tab:vibrational_modes_pCB}) shown in the solid blue line (using Eq.~\eqref{eq: Mollow_params}) as a function of dephasing rate (bottom x-axis) and temperature (top x-axis) whose correspondence is shown in Fig.~\ref{fig:DBT_spectra_dephasing}(b). The dashed blue line represents $8\Lambda_\alpha=1$. The solid red line depicts the normalized Mollow splitting (using Eq.~\eqref{eq: Mollow_params}) versus dephasing rate/temperature. The dotted vertical blue line marks $8\Lambda_\alpha=1.01$, to the left of which these quantities remain nearly constant. (b, c) Distinctively different characteristics of the Mollow triplets in ZPL and local phonon sidebands of a DBT molecule in pDCB (Table \ref{tab:vibrational_modes_pCB}) shown by temperature dependent variation of ratios of line broadening of the central and sidepeak ($R_\Gamma$) as well as their amplitude ratios ($R_{\rm A}$). In (b) the dashed colored lines represent the plots where the approximation $8\Lambda_\alpha = 1$ is used while the dashed black line represents $R_{\rm A} = 1$ below which the central peak will have lower amplitude than the sidepeaks.}
    \label{fig:ratios}
\end{figure}

In Fig.~\ref{fig:ratios}(a) we verify the validity of the condition $8|\Lambda_\alpha| = 1$ with respect to the dephasing rate which can be generalized to any quantum emitter. We find that as long as $T< 20$ K (or $\gamma_{\rm pd} < 200\gamma$) the approximation remains accurate exhibiting only a negligible difference. In Fig.~\ref{fig:ratios}(b) in black solid line we show the peak-broadening ratio $R_\Gamma$ of the Mollow triplets in ZPL as a function of temperature (dephasing rate) changing from $1:1.5$ to $2:1$, which is a marked departure from the standard Mollow triplet without pure dephasing. In Fig.~\ref{fig:ratios}(c) we show how the amplitude of Mollow triplets in the ZPL (black solid line) is significantly reduced with an increase in temperature compared to that of the local phonon sidebands. From Eq. (5) in the main text, we find that under finite dephasing, 
the central peak of the ZPL dephases at twice the rate of the sidepeaks. With increasing temperature, the spectral lines of the Mollow triplet not only result in equal amplitudes (marked by dotted black line), but at higher temperatures, the amplitude of the central peak may even become smaller than the sidepeaks since then the condition $8\Lambda_\alpha \approx 1 $ is no longer met as $8\Lambda_\alpha < 1$. The linewidth and amplitude ratios in the local phonon sidebands (shown in colored lines in Fig.~\ref{fig:ratios}) of DBT molecule on the other hand has different temperature dependent attributes compared to that in the ZPL as is shown in Fig.~\ref{fig:ratios}. Within the approximation of $8|\Lambda_\alpha| = 1$, and until $T \lesssim  5$ K, $R_\Gamma = 1:1$ and $R_{\rm A} = 2:1$ for all the modes in question and thus can be approximated to the case of zero dephasing. While for $T > 5$ K, $R_\Gamma$ increases while $R_{\rm A}$ decreases monotonically. 
\section{Analytical vs numerical calculations}
In this section we showcase the agreement between the numerical and analytical results derived in the strong driving limit pertaining to Case I and Case II.  
\begin{figure}[H]
    \centering
   \includegraphics[width = \textwidth]{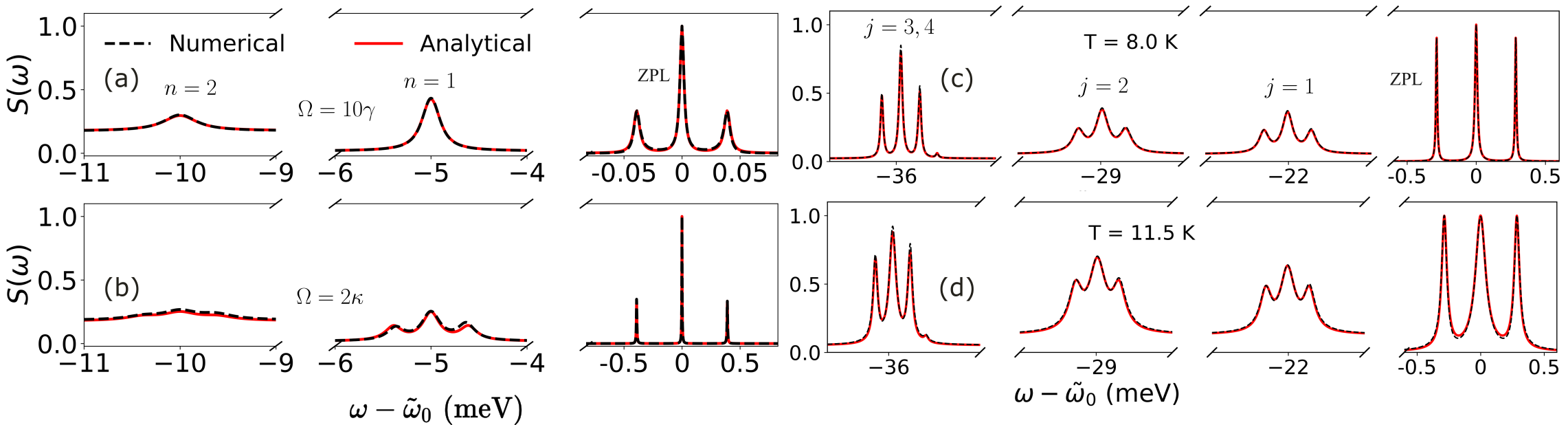}
    \caption{\small{(a) Broken x-axis plot showing agreement between the analytical (solid red line) and exact numerical (dashed dotted black line) spectra in meV frequency  scale for Case I where $\gamma_{\rm pd} = 0$ demonstrating ZPL and two phonon sideband. The parameters used are $\Omega = 10\gamma$, $\nu = 5$, $\eta = \nu/3$ meV and $\kappa = 0.2$ meV. (b)  Same parameters as (a) are used except that the driving frequency is $\Omega = 2\kappa$. (c) Normalized spectra of DBT molecule at temperature $T = 8$ K corresponding to $\gamma_{\rm pd} = 10\gamma$ driven by laser with Rabi frequency $\Omega = 2\kappa_2$ (Table \ref{tab:vibrational_modes_pCB}) calculated analytically (solid red line) and numerically (dashed black line) pertaining to Case II. The mode frequencies $\nu_i$, electron-phonon coupling $\eta_i$ and the phonon decay rate $\kappa_i$ are given in Table I. The local phonon sidebands are enhanced 60 times (24 times for superposition of mode 3 and 4) for better clarity on the linear scale. (d) The same for a DBT molecule at $T = 11.5$ K corresponding to $\gamma_{\rm pd} = 50\gamma$. The local phonon sidebands are enhanced half times than the ones in (c).}}
    \label{fig:main1}
\end{figure}

\begin{table}[h!]
\centering
\begin{tabular}{|c|c|c|c|c|c|c|c|c|c|c|}
\hline
$\bm{j}$ & 1 & 2 & 3 & 4 & 5 & 6 & 7 & 8 & 9 & 10 \\
\hline
$\bm{v_i}$ & 22.01 & 28.98 & 35.67 & 35.93 & 49.98 & 56.95 & 83.13 & 94.89 & 95.28 & 97.27 \\
\hline
$\bm{\eta_i}$ & 4.17 & 6.06 & 2.83 & 10.78 & 6.00 & 7.25 & 9.33 & 6.37 & 7.00 & 10.91 \\
\hline
$\bm{\kappa_i}$ & 0.129 & 0.156 & 0.035 & 0.037 & 0.054 & 0.060 & 0.016 & 0.028 & 0.033 & 0.060 \\
\hline
\end{tabular}
\caption{\small{Local vibrational mode parameters in meV for DBT embedded in \textit{para} dichlorobenzene (pDCB) crystal extracted from \cite{Zirkelbach2022High-resolution}.}}
\label{tab:vibrational_modes_pCB}
\end{table}

To assess the validity of the adiabatic approximation in the analytical formulation, we employ the root mean squared error (RMSE) to quantify the deviation between analytical and numerical results as a function of the normalized electron–phonon coupling $\eta/\nu$ and the Rabi frequency $\Omega/\eta$.This RMSE value is calculated by,
\begin{equation}
    \mbox{RMSE} = \sqrt{\frac{1}{T}\sum_{\tau = 1}^T\left(|g_{\rm ana}^{(1)}(\tau)| - |g_{\rm num}^{(1)}(\tau)|\right)^2 }
\end{equation}
\begin{figure}[H]
    \centering
    \includegraphics[width=0.5\linewidth]{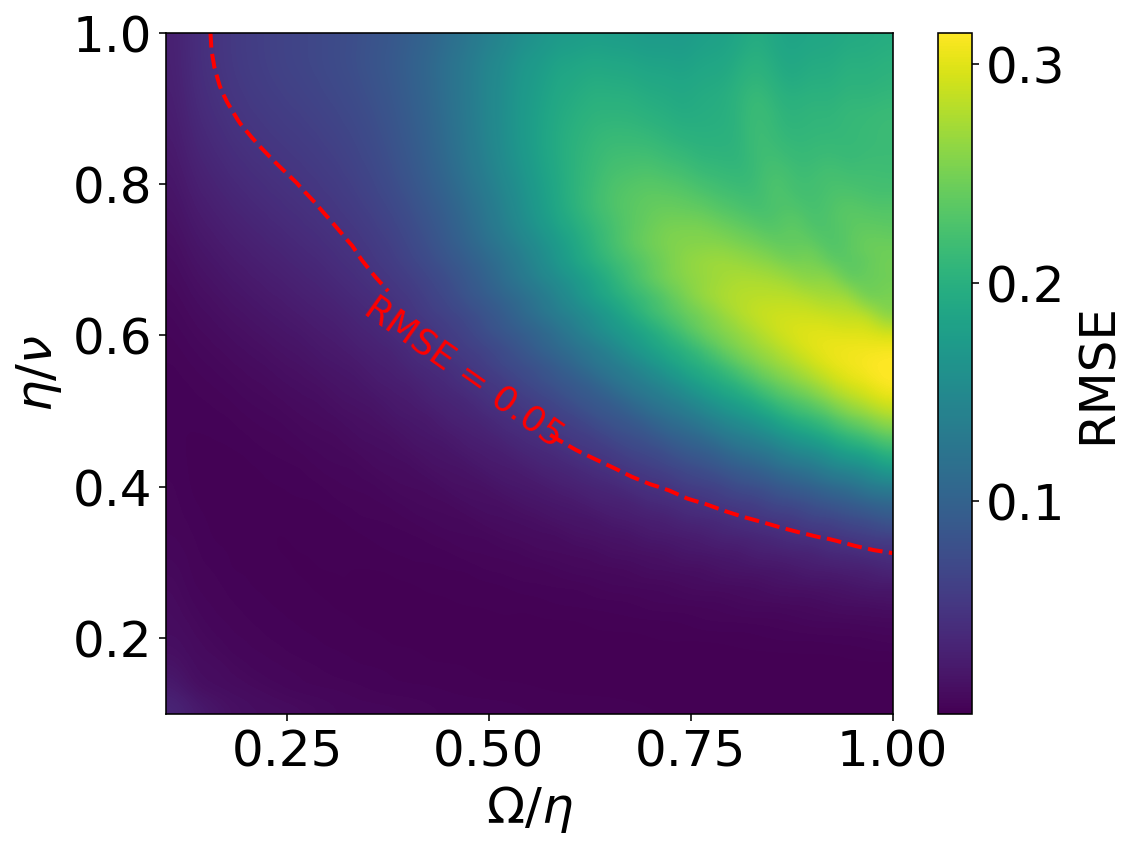}
    \caption{Root mean square error between numerical and analytical first order correlation function for a single phonon mode calculated upto 30 picoseconds with respect to the electron-phonon coupling rate $\eta$ and the Rabi frequency $\Omega$. The red dashed line corresponds to a cutoff defined by RMSE = 0.05. }
    \label{fig:error_color}
\end{figure}
The above figure is calculated up to $T = 30$ ps, which captures all the decay associated with the local phonon modes, as this is the region where the disagreement between the analytical and numerical results is most pronounced. This results in the color plot Fig.~\ref{fig:error_color} where the region at and below the red dashed line identifies the parameter regime defined by the coupling constant ($\eta$) and Rabi frequency ($\Omega$) where the analytical and numerical results coincide.

\section{Rabi frequency to laser power}
The intensity of the laser (Power/unit area) can be obtained from the time averaged value of the Poynting vector as $I = c\varepsilon_0E_0^2/2$, where $E_0$ is the amplitude of the laser field. The Rabi frequency tells us about the light-matter coupling between the laser and the quantum emitter given by $dE_0/\hbar$ where $d$ is the dipole moment of the quantum emitter. Thus we can obtain the laser intensity in terms of the Rabi frequency as follows,
\begin{equation}
    I = \frac{\varepsilon_0c}{2}\left(\frac{\hbar\Omega}{d}\right)^2 
\end{equation}
From the above equation we obtain the power as $P = I\cdot A$, where $A$ is the area of the laser spot. Now what remains to be calculated is the dipole moment which we do with the help of the decay rate of the quantum emitter that follows the relation $\gamma = \omega_0^3d^2/(3\pi\varepsilon_0\hbar c^3)$ from which we obtain,
\begin{equation}
    d = \sqrt{\frac{3 \varepsilon_0 \hbar\gamma \lambda^3}{8\pi^2}}
\end{equation}
where $\lambda$ is the wavelength of the emission from the quantum emitter.

\end{document}